\documentclass[showpacs,10pt,twocolumn,prb]{revtex4-1}

\usepackage{amsmath}
\usepackage{amssymb}
\usepackage{graphicx}
\usepackage{amssymb}
\usepackage{graphics}
\usepackage{epsfig}
\usepackage{color}

\begin{document}

\title{Insulating and metallic spin glass in K$_{x}$Fe$_{2-\delta-y}$Ni$_{y}$Se$_{2}$ (0.06 $\leq$ $y$ $\leq$ 1.44) single crystals}
\author{Hyejin Ryu,$^{1,2,\dag}$ Milinda Abeykoon,$^{1}$ Kefeng Wang,$^{1,\S}$ Hechang Lei,$^{1,\ddag}$  N. Lazarevic,$^{3}$ J. B. Warren,$^{4}$ E. S. Bozin,$^{1}$ Z. V. Popovic,$^{3}$ and C. Petrovic$^{1,2}$}
\affiliation{$^{1}$Condensed Matter Physics and Materials Science Department, Brookhaven
National Laboratory, Upton, New York 11973, USA\\
$^{2}$Department of Physics and Astronomy, Stony Brook University, Stony Brook, New York 11794-3800, USA\\
$^{3}$Center for Solid State Physics and New Materials, Institute of Physics Belgrade, University of Belgrade, Pregrevica 118, 11080 Belgrade, Serbia\\
$^{4}$Instrument Division, Brookhaven National Laboratory, Upton, New York 11973, USA}

\date{\today}

\begin{abstract}

We report electron doping effects by Ni in K$_{x}$Fe$_{2-\delta-y}$Ni$_{y}$Se$_{2}$ (0.06 $\leq$ $y$ $\leq$ 1.44) single crystal alloys. A rich ground state phase diagram is observed. Small amount of Ni ($\sim$ 4\%) suppressed superconductivity below 1.8 K, inducing insulating spin glass magnetic ground state for higher Ni content. With further Ni substitution, metallic resistivity is restored. For high Ni concentration in the lattice the unit cell symmetry is high symmetry $I4/mmm$ with no phase separation whereas both $I4/m + I4/mmm$ space groups were detected in the phase separated crystals when concentration of Ni $<$ Fe. The absence of superconductivity coincides with the absence of crystalline Fe vacancy order.

\end{abstract}

\maketitle

\section{Introduction}

The discovery of LaFeAsO$_{1-x}$F$_{x}$ has generated considerable attention due to unexpectedly high T$_{c}$'s of up to 52 K in isostructural Fe-based materials.\cite{Kamihara,ChenXH,ChenGF,RenZA,WenHH} Since then, several other types of iron-based superconductors have been discovered to include 122 structure (AFe$_{2}$As$_{2}$, A= Ca, Sr, Ba, and Eu),\cite{Rotter,ChenGF2} 111 structure (AFeAs, A=Li and Na),\cite{WangXC,Tapp} and 11 structure ($\alpha $-PbO type FeSe).\cite{Hsu} Recently, new family of iron-based superconductors A$_{x}$Fe$_{2-y}$Se$_{2}$ (A = K, Rb, Cs, and Tl) with maximum T$_{c}$ $\sim$ 33 K\cite{Guo,WangAF,Krzton-Maziopa,WangHD}  added not only to the materials variety but also to the complexity of the Fe-based superconductivity problem due to the intrinsic nanoscale phase separation and proximity to an insulating magnetic state with long range order.\cite{BaoW,RyanDH,WangZ,LiuY,ChenF,RicciA,LiW,YuanRH,LoucaD}

Perturbation of the ground state in Fe-based superconductors by chemical substitutions is rather instructive. Ba substition on KFe$_{2}$As$_{2}$ increases T$_{c}$ to 38 K,\cite{Rotter} whereas S doping in KFe$_{2}$Se$_{2}$ suppresses superconductivity.\cite{Hechang} Equally interesting is substitution of transition metal with unpaired 3d electrons and with similar ionic radius on Fe site. For instance, superconductivity is induced by Co and/or Ni doping in LaFeAsO, CaFeAsF, SrFe$_{2}$As$_{2}$, and BaFe$_{2}$As$_{2}$.\cite{Leithe-Jasper,Sefat,Matsuishi,Sefat1,Li} On the other hand, Co or Ni substitutions on Fe atomic positions in FeSe significantly suppresses superconductivity.\cite{Mizuguchi}

In this work, we have investigated structure, transport, magnetic, and optical properties of Ni substituted K$_{x}$Fe$_{2-\delta}$Se$_{2}$ single crystal series, where $\delta$ is Fe vacancy. Superconductivity is suppressed with much faster rate when compared to crystallographic phase separation. We observe rich ground state phase diagram where insulating magnetic spin glass gives way to metallic with further Ni concentration.

\section{Experiment}

Single crystals of K$_{x}$Fe$_{2-\delta-y}$Ni$_{y}$Se$_{2}$ used in this study were grown as described previously.\cite{Hechang,Lei1} The platelike crystals with typical size 5$\times$5$\times $2 mm$^{3}$ were grown. High energy synchrotron X-ray experiment at 300 K was conducted on X7B beamline of the National Synchrotron Light Source (NSLS) at Brookhaven National Laboratory (BNL). The setup utilized X-ray beam 0.5 mm $\times$ 0.5 mm in size with wavelength of 0.3196 ${\AA}$ (E = 38.7936 keV) configured with a focusing double crystal bent Laue monochomator, and Perkin-Elmer amorphous silicon image plate detector mounted perpendicular to the primary beam path. Finely pulverized samples were packed in cylindrical polyimide capillaries 1 mm in diameter and placed 377.81 mm away from the detector. Multiple scans were performed to a total exposure time of 240 s. The 2D diffraction data were integrated and converted to intensity versus 2$\theta$ using the software FIT2D.\cite{Hammersley} Structural refinements were carried out using GSAS modeling program\cite{Larson} operated by EXPGUI platform.\cite{Toby} The average stoichiometry was determined by energy-dispersive x-ray spectroscopy (EDX) in an JEOL JSM-6500 scanning electron microscope. Magnetization measurements, electric and thermal transport, and heat capacity were performed in a Quantum Design MPMS-XL5 and PPMS-9. Raman scattering measurements were performed on freshly cleaved samples using a JY T64000 Raman system in backscattering micro-Raman configuration. The 514.5 nm line of a mixed Ar$^+$/Kr$^+$ gas laser was used as an excitation source.\ The corresponding excitation power density was less than 0.2 kW/cm$^2$.\ Low temperature Raman measurements were performed using KONTI CryoVac continuous flow cryostat with 0.5 mm thick window.

The in-plane resistivity $\rho _{ab}(T)$ was measured using a four-probe configuration on cleaved rectangularly shaped single crystals with current flowing in the $ab$-plane of tetragonal structure. 
Thin Pt wires were attached to electrical contacts made of silver paste. Thermal transport properties were measured in Quantum Design PPMS-9 from 2 to 350 K using a one-heater two-thermometer method. The relative error was $\frac{\Delta \kappa}{\kappa}\sim$5$\%$ and $\frac{\Delta S}{S}\sim$5$\%$ based on Ni standard measured under identical conditions.

\section{Results and Discussion}

\begin{figure}
\centerline{\includegraphics[scale=0.40]{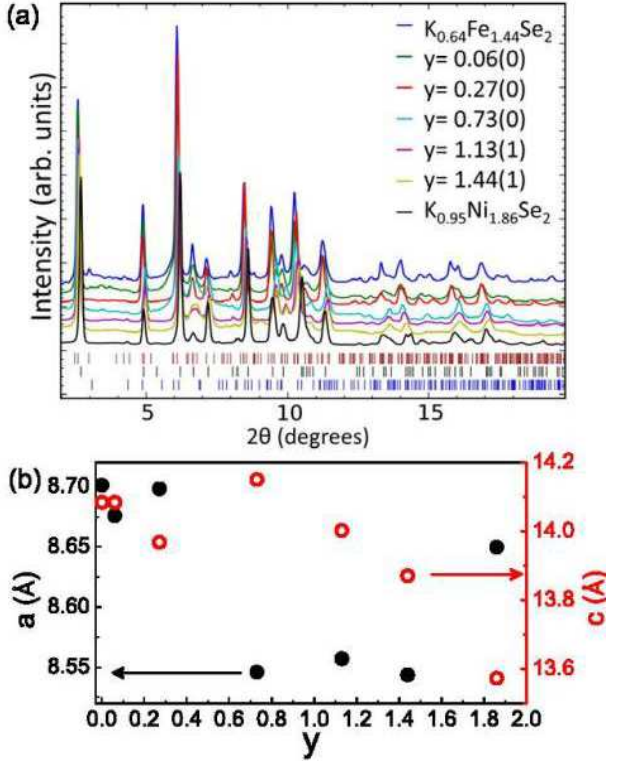}} \vspace*{-0.3cm}
\caption{(Color online) (a) High energy synchrotron X-ray diffraction data of K$_{x}$Fe$_{2-\delta-y}$Ni$_{y}$Se$_{2}$ series. Upper, middle, and lower vertical tick marks are for $I4/m$ phase, $I4/mmm$ phase, and SeO$_{2}$ reflections, respectively. (b) Lattice parameters as a function of Ni content y in K$_{x}$Fe$_{2-\delta-y}$Ni$_{y}$Se$_{2}$. Filled circles are for lattice parameter a and open circles are for lattice parameter c.}
\end{figure}

Obtained high energy synchrotron XRD results of K$_{x}$Fe$_{2-\delta-y}$Ni$_{y}$Se$_{2}$ series can be fitted very well with $I4/m$ and $I4/mmm$ space groups when y$\leq$0.73(0), while they are fitted by $I4/mmm$ space group only when y$\geq$1.13(1) (Fig. 1 (a)).  This implies $I4/m$ and $I4/mmm$ phases coexist when y$\leq$0.73(0). There is small amount of SeO$_{2}$ due to the oxidization. To make quantitative comparison of average structure in the alloy series (Fig. 1 (b)) only $I4/m$ space group is used for the refinements when y$\leq$0.73(0), since this results in the same goodness-of-fit when compared to the refinements using both $I4/m$ and $I4/mmm$ space groups. For y$\geq$1.13(1), \textit{I4/mmm} space group is used to determine the lattice parameters which are then converted into comparable numbers for $I4/m$ space group using the formula $I4/m = \sqrt{5}$ $I4/mmm$ for a-axis lattice parameters. Notice that there are considerable changes in lattice parameter $a$ around $y=0.73(0)$ and $y=1.44(1)$. On the other hand, the lattice parameter c starts to decrease when $y=0.73(0)$ as the Fe/Ni sites are filled with Ni. Nonmonotonic evolution of lattice parameters highlights complex crystal structure and bonding in K$_{x}$Fe$_{2-y}$Ni$_{y}$Se$_{2}$. Average atomic ratio of K, Fe, Ni and Se in crystals is shown in Table I. Defects and deficiency of Fe(Ni) stoichiometry is commonly found in AFeCh-122 compounds.\cite{Guo,Lei,Wang2} As the Ni ratio increases, deficiency of K and sum of Fe and Ni decreases, consistent with results on KNi$_{2}$Se$_{2}$ single crystals. \cite{Lei1}

\begin{figure}
\centerline{\includegraphics[scale=0.65]{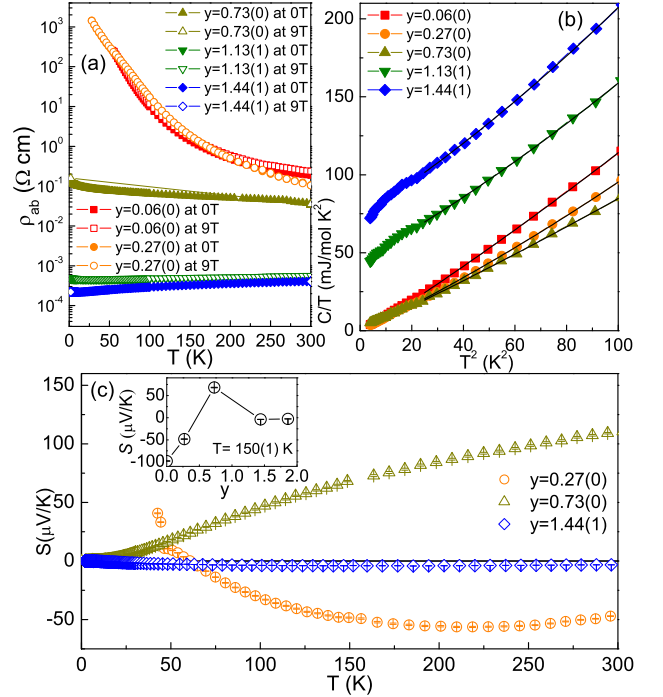}} \vspace*{-0.3cm}
\caption{(Color online) (a) Temperature dependence of the in-plane resistivity on K$_{x}$Fe$_{2-\delta-y}$Ni$_{y}$Se$_{2}$ series at zero and 9 T field. (b) The relation between C/T and T$^{2}$ for K$_{x}$Fe$_{2-\delta-y}$Ni$_{y}$Se$_{2}$ series at low temperature. The solid lines represent fits by the equation C/T=$\gamma$+$\beta_{3}$T$^{2}$+$\beta_{5}$T$^{4}$. (c) Temperature dependence of thermoelectric power S(T) for K$_{x}$Fe$_{2-\delta-y}$Ni$_{y}$Se$_{2}$ series. The inset shows the thermoelectric power at T=150 K for different Ni concentrations with adding results for y=0 and y=1.86(2) from Refs. 38 and 31.}
\end{figure}

\begin{table}[tbp]\centering%
\caption{Summary of measured stoichiometry of K$_{x}$Fe$_{2-\delta-y}$Ni$_{y}$Se$_{2}$ samples. The first column shows nominal values used in synthesis process.}%
\begin{tabular}{ccccc}
\hline\hline
K:Fe:Ni:Se & K & Fe & Ni & Se\\
\hline
1:1.8:0.2:2 & 0.73(0) & 1.37(1) & 0.06(0) & 2\\
1:1.4:0.6:2 & 0.87(2) & 1.15(1) & 0.27(0) & 2\\
1:1:1:2 & 0.84(0) & 0.85(0) & 0.73(0) & 2\\
1:0.6:1.4:2 & 0.83(0) & 0.56(0) & 1.13(1) & 2\\
1:0.2:1.8:2 & 0.82(0) & 0.26(1) & 1.44(1) & 2\\
\hline\hline
\end{tabular}%
\label{4}%
\end{table}%

K$_{0.8}$Fe$_{2}$Se$_{2}$ shows superconductivity below 30 K and metal to semiconductor transition at higher temperatures.\cite{Guo} As shown in Fig. 2 (a), 4.2 \% of Ni doping on K$_{0.8}$Fe$_{2}$Se$_{2}$ single crystal suppresses not only superconductivity but also conductivity and results in an insulating $\rho(T)$. However, as Ni substitutes for Fe further, conductivity increases up to the highest Ni concentration in  K$_{0.95}$Ni$_{1.86}$Se$_{2}$, consistent with previous study.\cite{Lei1}

\begin{table}[tbp]\centering%
\caption{Summary of $\gamma$ values and Debye temperatures in K$_{x}$Fe$_{2-\delta-y}$Ni$_{y}$Se$_{2}$}%
\begin{tabular}{ccc}
\hline\hline
y & $\gamma$(mJ mol$^{-1}$ K$^{-2}$) & $\Theta_{D}$(K)\\
\hline
0.06(0) & 0.6(4) & 210(3)\\
0.27(0)& 0.3(2) & 230(6)\\
0.73(0)& 0.2(1) & 232(2)\\
1.13(1)& 45(7) & 218(3)\\
1.44(1)& 72(9) & 208(5)\\
\hline\hline
\end{tabular}%
\label{4}%
\end{table}%

Relation between C/T and T$^{2}$ also shows insulator to metal transition as Ni concentration increases, as shown in Fig. 2 (b). The electronic specific heat and Debye temperature are obtained by the fitting on C/T-T$^{2}$ curves from 5 K to 10 K region using the formula C/T=$\gamma$+$\beta_{3}$T$^{2}$+$\beta_{5}$T$^{4}$. The Debye temperatures are estimated by the formula $\Theta_{D}=(12\pi^{4}NR/5\beta)^{1/3}$, where N is the atomic number in the chemical formula and R is the gas constant. The obtained $\gamma$ values and Debye temperatures $\Theta_{D}$ for different Ni concentrations are listed in the Table II. All samples in K$_{x}$Fe$_{2-\delta-y}$Ni$_{y}$Se$_{2}$ series have similar $\Theta_{D}$ values which reflects no significant changes in atomic weight, structure and bonding. In addition, $\gamma$ values are negligible for $y\leq0.73(0)$ suggesting minute density of states at the Fermi level as expected for an insulator. Larger $\gamma$ values for $y>0.73(0)$ region reflect rapid pileup of the density of states $N(E_{F})$ in the metallic region and possible heavy fermion-like behavior.\cite{Lei1,Neilson}

Temperature dependence of thermoelectric power S(T) for K$_{x}$Fe$_{2-\delta-y}$Ni$_{y}$Se$_{2}$ series is shown in the main part whereas S(y) at 150 K is presented in the inset of Fig. 2(c). Large Fermi surface changes are evident around y=0.73(0); this coincides with the notable lattice parameter change in XRD results. The changes are related to the metal insulator transition. It is interesting that the dominant carriers for end members of K$_{x}$Fe$_{2-\delta-y}$Ni$_{y}$Se$_{2}$ (y=0 and y=1.86(2))\cite{Lei1,Kefeng} are electrons at high temperature whereas they are holes for samples in the middle of the series.

\begin{figure}
\centerline{\includegraphics[scale=0.40]{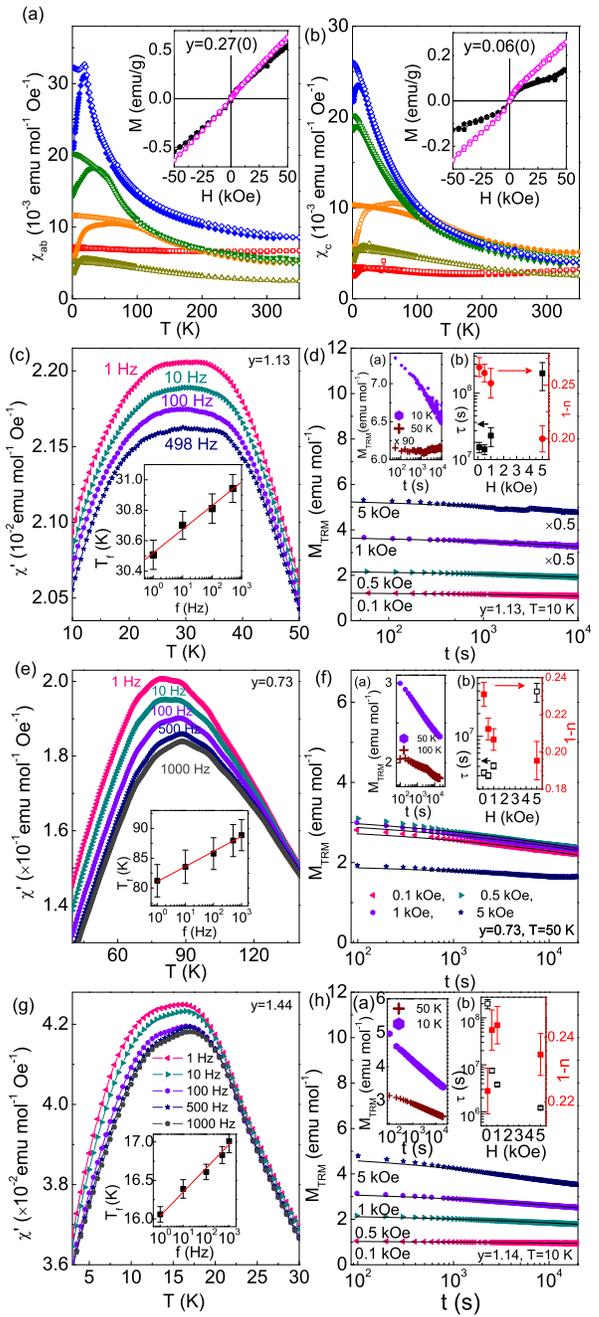}} \vspace*{-0.3cm}
\caption{(color online) Temperature dependence dc magnetic susceptibilities for K$_{x}$Fe$_{2-\delta-y}$Ni$_{y}$Se$_{2}$ series for (a) H$\bot$c and for (b) H$\|$c at H=1 kOe in ZFC and FC. Inset figures of (a) and (b) are M-H loops for H$\bot$c and H$\|$c, respectively at 1.8 K (filled pentagon) and 300 K (open pentagon). (c) Temperature dependence of $\chi'(T)$ measured at several fixed frequencies for y=1.13(1) of K$_{x}$Fe$_{2-\delta-y}$Ni$_{y}$Se$_{2}$. Inset is the frequency dependence of $T_{f}$ with the linear fitting (solid line). (d) TRM versus time for y=1.13(1) of K$_{x}$Fe$_{2-\delta-y}$Ni$_{y}$Se$_{2}$ at 10 K and $t_{w}=100s$ with different dc field with fitting (solid lines). Inset (a) is $M_{TRM}$ vs. t at 10 K and 50 K at H = 1 kOe and $t_{w}=100s$. Inset (b) is H-field dependence $\tau(s)$ (filled square) and 1-$n$ (filled circle).(e-h) Similar data for y=0.73 and 1.44 }
\end{figure}

Temperature dependent anisotropic magnetization for K$_{x}$Fe$_{2-\delta-y}$Ni$_{y}$Se$_{2}$ series is shown in Fig. 3(a) and (b). A pronounced irreversible behaviors between zero-field-cooling (ZFC) and field-cooling (FC) curves below 50 K is observed. The irreversibility implies a magnetic spin glass where the spins are locked or frozen into random orientations below a characteristic temperature T$_{f}$. Similar behavior has been reported in TlFe$_{2-x}$Se$_{2}$, KFeCuS$_{2}$, and K$_{x}$Fe$_{2-\delta}$S$_{2}$.\cite{Ying,Oledzka,Lei} M-H loops in insets of Fig. 3(a) and (b) also support glassy nature of the transition by presenting nearly linear field dependence with no hysteresis at 300 K or s-shape loop at 1.8 K.\cite{Ying} Fig. 3(c) shows the frequency dependent peak of the real part in ac susceptibility $\chi'(T)$. As the frequency increases, peak position moves to the higher temperature while magnitude decreases, another hallmark of the typical spin glass behavior.\cite{Mydosh} Relation between T$_{f}$ and frequency can be fitted by K=$\Delta$T$_{f}$/(T$_{f}$$\Delta$log$f$), and the obtained K value is 0.0050(2). Similarly, K value for y=0.73 is 0.030(3) and for y=1.44 is 0.020(2) [Fig. 3(e-h)]. This is in agreement with the values (0.0045 $\leq$ K $\leq$ 0.08) for a canonical spin glass.\cite{Mydosh} Thermoremanent magnetization (TRM) is shown in Fig.3 (d). The sample was cooled down from 100 K (above T$_{f}$) to 10 K (below T$_{f}$) in different magnetic fields, and kept there for $t_{w}=100s$. Then, magnetic field was switched off and magnetization decay M$_{TRM}$(t) was measured. At T = 10 K, M$_{TRM}$(t) shows slow decay, so M$_{TRM}$(t) has non-zero values even after several hours. The slow decay of M$_{TRM}$(t) is another typical property of magnetic spin glass.\cite{Mydosh} On the other hand, at T = 50 K (above T$_{f}$), M$_{TRM}$(t) decays rapidly in a short time and stays nearly constant, when compared to the data at T = 10 K as shown in Fig. 3(d) inset (a). M$_{TRM}$(t) decay in spin glass system is commonly explained by stretched exponential function, M$_{TRM}$(t) = $M_{0}exp[-(t/\tau)^{1-n}]$, where $M_{0}$, $\tau$, and 1-$n$ are the glassy component, the relaxation characteristic time, and the critical exponent, respectively. We observe that $\tau$ is significantly increased while 1-$n$ is slightly decreased in magnetic field (Fig. 3 (d) inset (b)). The value of 1-$n$ is close to 1/3, as expected for a magnetic spin glass.\cite{Campbell,Chu}

\begin{figure}
\centerline{\includegraphics[scale=0.5]{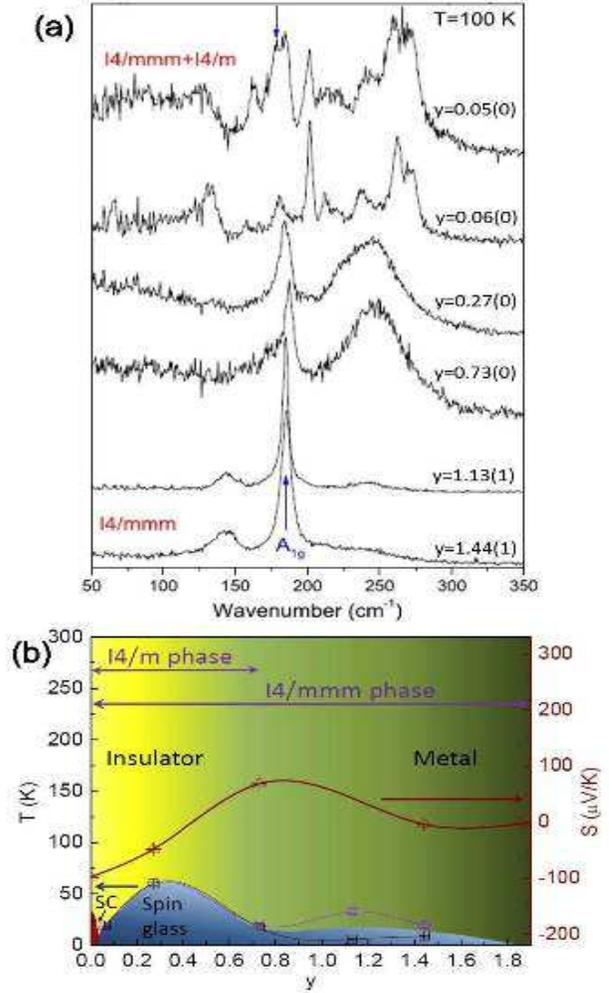}} \vspace*{-0.3cm}
\caption{(color online) (a) Raman scattering spectra of K$_{x}$Fe$_{2-\delta-y}$Ni$_{y}$Se$_{2}$ single crystal series measured from the $ab$ plane at 100 K. (b) Magnetic, transport, and structural phase diagram. The purple circles are H$\bot$c and the black squares are for H$\|$c, respectively.}
\end{figure}

According to symmetry considerations one can expect four Raman-active phonons for the $I4/mmm$ phase (A$_{1g}$, B$_{1g}$ and 2E$_{g}$).\cite{Lazarevic1} However, ordering of the vacancies, as reported for the K$_x$Fe$_{2-\delta}$Se$_2$,\cite{Lazarevic2} locally reduces the symmetry to $I4/m$. This leads to substantial increase in the number of Raman modes. Symmetry analysis predicts total of 27 Raman modes (9A$_{g}$, 9B$_{g}$ and 9E$_{g}$) originating from the vibrations within the $I4/m$ phase. When Raman spectra are measured from the $ab$ plane of the sample, only two Raman modes can be seen for the $I4/mmm$ phase (A$_{1g}$+B$_{1g}$) and 18 for the $I4/m$ phase (9A$_{g}$+9B$_{g}$).

Fig. 4 (a) shows Raman scattering spectra measured at 100 K from the $ab$ plane of K$_{x}$Fe$_{2-\delta-y}$Ni$_{y}$Se$_{2}$ single crystal series. For the high concentration of Ni ($y=$1.44(1)) only two modes can be observed. These modes were previously assigned as A$_{1g}$ (185 cm$^{-1}$) and B$_{1g}$ (141 cm$^{-1}$) modes. We notice traces of an additional structure around 248 cm$^{-1}$ for the $y=1.13(1)$ crystal. The structure is present for all investigated samples $y\leq1.13(1)$ but it is highly pronounced for the $y=0.73(0)$ and $y=0.27(0)$ samples. The origin of this structure is most likely related to crystalline disorder. Disorder breaks the conservation of the momentum during the Raman scattering process enabling contributions of finite wavevector phonons to Raman spectra. Another possibility is an appearance of new high symmetry phase. However this finding is not supported by XRD measurements. For the low concentrations of Ni the structure at around 248 cm$^{-1}$ vanishes and the large number of vibrations of the $I4/m$ phase are observed, suggesting vibrations from vacancy ordered domains in the crystal. The A$_{1g}$ mode (marked by arrow in Fig. 4 (a)), which represent the vibration of selenium ions in the $I4/mmm$ phase, persist for all Ni concentrations. This shows the presence of the $I4/mmm$ phase in all samples.

Our main results are summarized on Fig. 4 (b) phase diagram. As shown in the lower left corner of the phase diagram, superconducting (SC) phase disappears rapidly by y=0.06(1). The $I4/mmm$ space group is found for all K$_{x}$Fe$_{2-\delta-y}$Ni$_{y}$Se$_{2}$ series whereas crystalline disordered $I4/m$ space group persists up to $y=0.73$. Hence, Fe-based high temperature superconductivity in K$_{x}$Fe$_{2-\delta-y}$Ni$_{y}$Se$_{2}$ does vanish before crystalline superstructure of Fe vacancies (crystalline ordered $I4/m$ phase) disappears when \textit{y} is increased from 0. We note that in high pressure studies superconductivity vanishes simultaneously with $I4/m$ superstructure peak.\cite{Guo1} High degree of crystalline disorder in $I4/mmm$ and in $I4/m$ phase results in insulating or bad metal magnetic glass state that borders superconducting region, similar to copper oxides.\cite{RaicevicI,ShiX}
In the insulating region of K$_{x}$Fe$_{2-\delta-y}$Ni$_{y}$Se$_{2}$ single crystal alloys ground state phase diagram (Fig. 4(b)), freezing temperatures T$_{f}$
of the magnetic spin glass are higher ($\sim$ 60 K) when compared to metallic regions ($\sim$ 20 K) (y$>$1.13(1)).

The mechanism of the nonmetallic states in proximity to K$_{x}$Fe$_{2-\delta}$Se$_{2}$ is of the great importance for the understanding of superconductivity.\cite{YinZP,DaiP} Intimate nanoscale mix of superconducting and
insulating magnetic regions may also add states at interfaces.\cite{YanYJ,MukherjeeS} This makes interpretation of many, and in particular bulk measurements difficult. In K$_{x}$Fe$_{2-\delta}$Se$_{2}$ nanoscale phase separation is found below T$_{s}$ = 560 K,\cite{BaoW} hence majority of conductivity changes at temperature below $T_{s}$ should come from the metallic regions. This is supported by recent angle-resolved photoemission results where
orbital-selective Mott transition in K$_{x}$Fe$_{2-\delta}$Se$_{2}$ was observed above 150 K.\cite{YiM} This temperature corresponds to metal - insulator
crossover in bulk measurements, suggesting that conductivity changes in bulk measurements may not be simply
due to the ratio of metallic and insulating regions in the crystal.\cite{Guo,Shoemaker} Though only metallic nanoscale regions contribute to thermopower and metallic heat capacity (Fig. 2(b,c)), the absolute values of resistivity and magnetization reflect the contribution of both ($I4/mmm$) and insulating parts of the crystal ($I4/m$). Assuming that Ni substitutes Fe in both space group, small Ni substitution therefore is likely to have strong effect on states associated with itinerant $d_{xz}/d_{yz}$ orbitals, perhaps via
localization effect in an orbital-selective Mott localization scenario.\cite{YuRong,YuRong2} Further Ni substitution and disorder might enhance conductivity by raising chemical potential and enlarging electron pockets.\cite{BerlijnT,CracoL,LuF} This is in agreement with our phase diagram.

\section{Conclusion}

We have investigated transport, magnetic and structure changes in K$_{x}$Fe$_{2-\delta-y}$Ni$_{y}$Se$_{2}$ single crystal series. Small amount of Ni doping $y=0.06$ suppressed Fe-based high temperature superconductivity. The suppression of superconductivity is more sensitive to Ni substitution than crystalline superstructure of Fe vacancies. Further Ni substitution results in insulating and bad metal magnetic spin glass ground state.
However, when Ni concentration in the lattice is higher than Fe, metallic ground state with relatively large density of states at the Fermi level emerges. Similar to copper oxides, insulating/bad metal spin glass is found in proximity to superconducting state. In Ni substituted in K$_{x}$Fe$_{2-\delta}$Se$_{2}$ the spin glass state covers nearly all $y$ values, from superconductivity up to the paramagnetic metal K$_{x}$Ni$_{2-\delta}$Se$_{2}$.
\begin{acknowledgements}

Work at Brookhaven is supported by the U.S. DOE under Contract No. DE-AC02- 98CH10886 and in part
by the Center for Emergent Superconductivity, an Energy Frontier Research Center funded by the U.S. DOE, Office for Basic Energy Science (K. W and C. P). This work was also supported by the
Serbian Ministry of Education, Science and Technological Development under Projects ON171032 and III45018.

\end{acknowledgements}

\dag Present address: Advanced Light Source, E. O. Lawrence Berkeley National Laboratory, Berkeley, California 94720, USA

\S Present address: CNAM, Department of Physics, University of Maryland, College Park, Maryland 20742, USA

\ddag\ Present address: Frontier Research Center, Tokyo Institute of Technology, 4259 Nagatsuta, Midori, Yokohama 226-8503, Japan.

\end{document}